\documentclass[10pt,conference]{IEEEtran}

\usepackage{cite}
\usepackage{amsmath,amssymb,amsfonts}
\usepackage{graphicx}
\usepackage{textcomp}
\usepackage{xcolor}
\usepackage[hyphens]{url}
\usepackage{fancyhdr}
\usepackage{hyperref}

\usepackage{xspace}          
\usepackage{booktabs}        
\usepackage{makecell}        
\usepackage{multirow}        
\usepackage{array}           
\usepackage{subcaption}      
\usepackage{siunitx}         
\usepackage[capitalize,noabbrev]{cleveref}  
\crefname{algocf}{Algorithm}{Algorithms}
\Crefname{algocf}{Algorithm}{Algorithms}
\usepackage{threeparttable}  
\usepackage[ruled,linesnumbered,vlined]{algorithm2e}
\usepackage{setspace}
\usepackage{etoolbox}

\DeclareMathOperator*{\argmin}{arg\,min}
\newcommand{\Qround}[2]{\mathrm{round}(#1,#2)}
\newcommand{\Rset}{R}
\newcommand{\Nbr}[1]{N_4(#1)}

\hypersetup{
  colorlinks=true,
  linkcolor=blue!60!black,
  citecolor=blue!60!black,
  urlcolor=blue!60!black,
}

\title{Heterogeneity-Aware Microscaling for Efficient Low-Bit LLM Inference}

\author{%
  \IEEEauthorblockN{%
    Junyi Luo\IEEEauthorrefmark{1},
    Xinting Jiang\IEEEauthorrefmark{1},
    Tai-Hao Wen\IEEEauthorrefmark{2},
    Ruichen Qi\IEEEauthorrefmark{1},
    Minxing Chu\IEEEauthorrefmark{1},
    Hongyi Wu\IEEEauthorrefmark{2}, \\
    Gregory Kielian\IEEEauthorrefmark{3},
    Ben Laurie\IEEEauthorrefmark{3},
    Qirui Zhang\IEEEauthorrefmark{2},
    Quan Cheng\IEEEauthorrefmark{1},
    Dennis Sylvester\IEEEauthorrefmark{2},
    Mehdi Saligane\IEEEauthorrefmark{1}%
  }
  \IEEEauthorblockA{%
    \IEEEauthorrefmark{1}Brown University\quad
    \IEEEauthorrefmark{2}University of Michigan\quad
    \IEEEauthorrefmark{3}Google\\[2pt]
    \{junyi\_luo, mehdi\_saligane\}@brown.edu%
  }%
}

\newcommand{\AdaMX}{AdaMX\xspace}
\newcommand{\AdaMXSixteen}{AdaMX-16\xspace}
\newcommand{\AdaMXThirtyTwo}{AdaMX-32\xspace}

\newcommand{\WFmt}{E4Mt2T2\xspace}
\newcommand{\AFmt}{E4M1Mt2N1\xspace}

\newcommand{\Efour}{\text{E4}\xspace}

\newcommand{\Eeight}{\text{E8M0}\xspace}
\newcommand{\Mttwo}{\text{Mt2}\xspace}
\newcommand{\Ttwo}{\text{T2}\xspace}
\newcommand{\None}{\text{N1}\xspace}
\newcommand{\Mone}{\text{M1}\xspace}
\newcommand{\AScale}{\text{E4M1}\xspace}  

\newcommand{\MMXFP}{M2XFP\xspace}     
\newcommand{\MXplus}{MX+\xspace}      
\newcommand{\MXFPfour}{MXFP4\xspace}
\newcommand{\MXINTfour}{MXINT4\xspace}
\newcommand{\NVFPfour}{NVFP4\xspace}
\newcommand{\FPfour}{FP4\xspace}
\newcommand{\INTfour}{INT4\xspace}
\newcommand{\FPsix}{FP6\xspace}

\newcommand{\EBW}{EBW\xspace}

\let\oldsubsubsection\subsubsection
\renewcommand{\subsubsection}[1]{\oldsubsubsection{\normalfont\bfseries #1}}

\SetAlFnt{\small}
\SetAlCapNameFnt{\small}
\SetAlCapFnt{\small}
\SetAlgoNlRelativeSize{-0.5} 
\IncMargin{3pt}

\SetKwInput{KwIn}{Input}
\SetKwInput{KwOut}{Output}
\SetKw{KwRet}{return}

\begin{document}
\maketitle
\pagestyle{plain}


\begin{abstract}

Microscaling (MX) is now the standard for low-bit large language model (LLM) inference.
Its 4-bit form \MXFPfour still loses substantial accuracy, because existing MX formats fix either the element format or the precision-recovery scheme across blocks, and thus capture only limited quantization heterogeneity.
Quantization heterogeneity appears at two levels: 1) across blocks, the preferred element format and precision-recovery scheme vary; 2) across operands, weights and activations require different encoding.
We introduce \AdaMX (Adaptive Microscaling), a heterogeneity-aware format and accelerator.
It selects the precision-recovery scheme per block and the representation per operand, at no increase in equivalent bit width (\EBW).
One design covers two block sizes, giving a higher-accuracy operating point and a lower-\EBW operating point that saves storage.
We implement a 22nm FD-SOI AI accelerator prototype with the proposed decoder, computing unit, and quantization logic.
Against an otherwise identical \MXFPfour accelerator with FP4-only multipliers, \AdaMX adds about $1\%$ system energy.
At the lower-\EBW point, \AdaMX stays more accurate than the baseline while lowering both memory footprint and energy.
Across LLMs from 3B to 70B, \AdaMX removes 83\% of the \MXFPfour accuracy loss on commonsense and 82\% on MMLU, and 43\% and 27\% of the \NVFPfour loss.
\AdaMX also generalizes to multimodal models.
On Gemma-4 12B, it leads \MXFPfour on all four vision-language benchmarks and keeps up to $96\%$ of FP16 accuracy.

\end{abstract}

\section{Introduction}
\label{sec:intro}

LLM inference is memory-bound.
Recent models reach hundreds of billions of parameters~\cite{gpt5, claude, deepseekv4}, so reading their weights from off-chip memory dominates both latency and energy, with each off-chip access far costlier than an arithmetic operation~\cite{horowitz2014}.
Low-bit quantization shrinks the bits moved per weight and activation and is the main lever for efficient LLM inference; it also lets a fixed memory budget hold a larger model.

The Microscaling (MX) format has become the industry standard for this purpose.
It is standardized by the Open Compute Project~\cite{ocpmx} and natively supported in recent accelerators such as NVIDIA Blackwell~\cite{nvfp4}, AMD Instinct MI350~\cite{amd_mi350}, Microsoft Maia 100~\cite{xu2024maia100}, and Meta MTIA~\cite{meta2026mtia}.
MX partitions a tensor into small blocks and gives each block a single \Eeight scale, a power-of-two format with 8~exponent (E) bits and 0~mantissa (M) bits.
Its 4-bit form \MXFPfour saves the most memory, but at this width the power-of-two scale is too coarse, and \MXFPfour degrades sharply in both perplexity (PPL) and downstream accuracy~\cite{rouhani2023mx,fouroversix}.

\begin{figure}[t]
  \centering
  \includegraphics[width=\columnwidth]{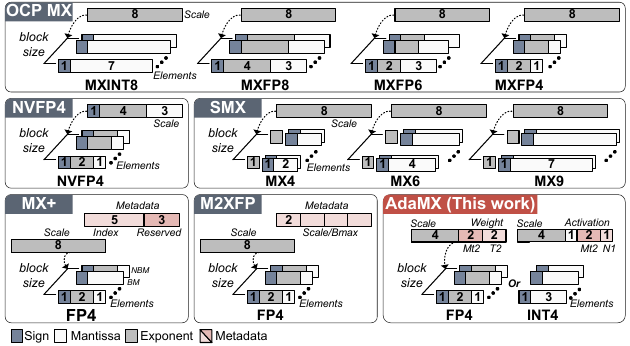}
  \caption{Bit layouts of MX-family formats. \AdaMX repartitions the 8-bit \Eeight scale into a narrower scale and per-block metadata: an \Efour scale with type routing for weights, and an \AScale scale with a lossless \FPsix block-max extension for activations.}
  \label{fig:figure1_data_formats}
\end{figure}

Recent formats narrow this gap in two ways (\cref{fig:figure1_data_formats}).
Some make the scale finer, often over smaller blocks: \NVFPfour pairs a smaller group with an FP8 scale~\cite{nvfp4}, and shared microexponents (SMX) add a sub-block exponent~\cite{smx}.
Others keep the block scale but attach per-block \emph{metadata}, a few extra bits that the decoder reads to recover the precision lost to the coarse scale, as in \MXplus~\cite{mxplus} and \MMXFP~\cite{m2xfp}.
Both recover accuracy, but each uses extra storage that raises the equivalent bit width (\EBW).

These formats share an implicit assumption that a single element format and recovery scheme are suitable for every block of an operand.
Our analysis shows that this assumption fails at two levels: the preferred format and recovery scheme vary across blocks, while weights and activations require different encoding strategies (\cref{sec:motivation}).
Meanwhile, the MX exponent field is wider than required by LLM blocks, leaving unused bits that can encode heterogeneity-aware metadata without increasing \EBW.

\begin{table*}[t]
\centering
\caption{Comparison of microscaling and metadata-augmented formats.
  Metadata granularity is stated where it differs from the scale granularity.
  }
\label{tab:prior}
\small
\renewcommand{\arraystretch}{1}
\setlength{\tabcolsep}{4pt}
\begin{tabular*}{0.93\textwidth}{@{\extracolsep{\fill}}lcccl@{}}
\Xhline{1.2pt}
\textbf{Design}
  & \textbf{Gran.} & \textbf{Scale Fmt}
  & \textbf{Element Fmt}
  & \textbf{Metadata} \\
\hline
\MXFPfour~\cite{ocpmx}
  & Blk-32 & \Eeight
  & FP4
  & -- \\
\NVFPfour~\cite{nvfp4}
  & Blk-16 & E4M3
  & FP4
  & -- \\
SMX~\cite{smx}
  & Blk-16 & \Eeight
  & INT4
  & 1-bit exponent per pair \\
BBAL~\cite{bbal}
  & Blk-32 & E5M0
  & INT4 (BFP)
  & 1-bit direction flag per element \\
MicroScopiQ~\cite{microscopiq}
  & Blk-128 & \Eeight
  & FP4 + INT4
  & outlier index + scale \\
\MMXFP~\cite{m2xfp}
  & Blk-32 & \Eeight
  & FP4
  & 2-bit sub-block ratio or block maximum \\
\MXplus~\cite{mxplus}
  & Blk-32 & \Eeight
  & FP4
  & 5-bit index + 3-bit reserved \\
BlockDialect~\cite{blockdialect}
  & Blk-32 & \Eeight
  & FP4 (dialect)
  & format index \\
\hline
\AdaMX~(weight)
  & Blk-16/32 & \Efour + per-channel bias
  & FP4\,/\,INT4
  & 2-bit type selector + 2-bit payload \\
\AdaMX~(activation)
  & Blk-16/32 & \AScale + per-token bias
  & FP4
  & 2-bit payload + 1-bit neighborhood selector \\
\Xhline{1.2pt}
\end{tabular*}
\end{table*}

This paper makes four contributions:
\begin{itemize}
  \item \textbf{Quantization heterogeneity in microscaling.}
  We characterize multi-level quantization heterogeneity in low-bit MX.
  Within a weight tensor, the preferred element format and precision-recovery mechanism vary independently across blocks; across operands, weights and activations require different adaptation: weights are static and admit an offline per-block search, while activations are produced online and must be encoded in a single pass.

  \item \textbf{The \AdaMX format.}
  We introduce \AdaMX, an operand-specialized microscaling family that adapts to this heterogeneity.
  On weights, a 2-bit selector routes each block among four combinations of element format (\FPfour/\INTfour) and recovery scheme (scale refinement or block-maximum extension), chosen by an offline search.
  On activations, encoded online, we systematically analyze scale rounding and derive a rounding rule co-designed with a lossless block-maximum extension, all without calibration.

  \item \textbf{A unified accelerator.}
  We design a 22-nm FD-SOI accelerator with one datapath for both operands: a \Ttwo-routed MAC that covers all four format-enhancement modes with a shared multiplier and a lightweight correction sidepath, and a streaming engine that encodes activations online.
  The same design serves both block sizes and adds at most about 1\% end-to-end energy over the simplest equal-throughput \MXFPfour baseline.

  \item \textbf{Evaluation.}
  Across LLMs from 3B to 70B and the Gemma-4 12B multimodal model, \AdaMX cuts the accuracy loss of \MXFPfour by 83\% and of \NVFPfour by 43\%, closing 63 to 72\% of the \MXFPfour perplexity gap to FP16.
\end{itemize}


\section{Background and Motivation}
\label{sec:background}

\subsection{Microscaling and Its 4-bit Formats}
\label{sec:bg:mx}

Post-training quantization maps high-precision tensors to low-bit representations, cutting the memory and compute of LLM inference~\cite{gptq,awq,smoothquant,quarot,quipsharp,atom,omniquant,zeroquant,llmint8,spqr,aqlm,squeezellm,tender}.
Block-wise quantization, which shares one scale across 16--128 elements, has become the prevailing choice~\cite{ocpmx,mant,ant,nvfp4,msfp,mxplus,m2xfp,nxfp}.
MX is widely studied and deployed~\cite{fouroversix,sageattention,m2xfp,mxplus,mx_cim_vlsi2025,cuyckens2025efficient,mx_multimode_gain_cell,isscc26_mxfp6}.
For a block $\{x_i\}$ and an element format whose largest power of two is~$P$, the shared exponent is $S = 2^{\lfloor \log_2(\max_i |x_i|/P)\rfloor}$~\cite{rouhani2023mx}.
Alternative scale-rounding rules exist~\cite{mishra2025mxfp8,yang2025arith,fouroversix}; \cref{sec:s1_encoding} evaluates the effect of this rounding choice.

Beyond the shared scale, the element type provides another design dimension in 4-bit MX formats.
\label{sec:bg:fpint}
\FPfour{} (E2M1) uses a 1-bit sign, a 2-bit exponent, and a 1-bit mantissa, forming a non-uniform grid that is denser near zero and better captures values spanning a wide dynamic range.
In contrast, \INTfour{} uses a 1-bit sign and a 3-bit magnitude, providing a uniform grid with finer resolution over a fixed range.
Chen et al.\ systematically analyze the accuracy trade-off between INT and FP under fine-grained MX quantization and find that \FPfour{} generally outperforms \INTfour{} at 4-bit precision~\cite{chen2025intvsfp}.
We revisit this tensor-level conclusion at the block level in \cref{sec:motivation}.

\subsection{Metadata-Augmented MX Formats}
\label{sec:bg:prior}

To recover the accuracy \MXFPfour loses to its coarse scale, recent designs augment each MX block with metadata~\cite{mxplus,m2xfp,microscopiq,blockdialect}.
\cref{tab:prior} compares them along two axes: the granularity of the metadata (element, subgroup, or block) and its content (extra mantissa bits, extra exponent bits, or a format index).
Two techniques are commonly used.
\textbf{Block-maximum mantissa extension} adds mantissa bits to the block's largest element, which dominates \MXFPfour error~\cite{mxplus}.
Because this element's exponent is fixed at the format's maximum, its exponent bits carry no information and can be reused as mantissa bits, promoting it from \FPfour to FP6 at no extra block storage.
\MXplus records the element's position with a 5-bit index, while \MMXFP clamps the FP6 result to a 4-value neighborhood with 2 bits~\cite{m2xfp}.
\textbf{Scale refinement} instead adds mantissa bits to the shared scale, thereby reducing quantization error across all block elements.
Because the \Eeight scale jumps in factors of two, up to half of a block's range goes unused when its maximum falls between two bins; \NVFPfour replaces \Eeight with an E4M3 scale, adding 3 mantissa bits~\cite{nvfp4}, and \MMXFP adjusts the sub-block scale with 2 bits~\cite{m2xfp}.
\Cref{fig:t2_axes} illustrates both mechanisms, and \cref{sec:design} details how \AdaMX applies them.

\subsection{The Heterogeneity Opportunity}
\label{sec:motivation}

Existing MX formats treat quantization as homogeneous.
We analyze the weights and activations of four LLMs (Llama-3.1-8B~\cite{llama3}, Qwen2.5-3B and Qwen2.5-14B~\cite{qwen25}, Gemma-2-9B~\cite{gemma2}) and identify two forms of heterogeneity and show that the MX exponent field provides sufficient capacity to encode them.

\begin{figure}[t]
  \centering
  \includegraphics[width=\linewidth]{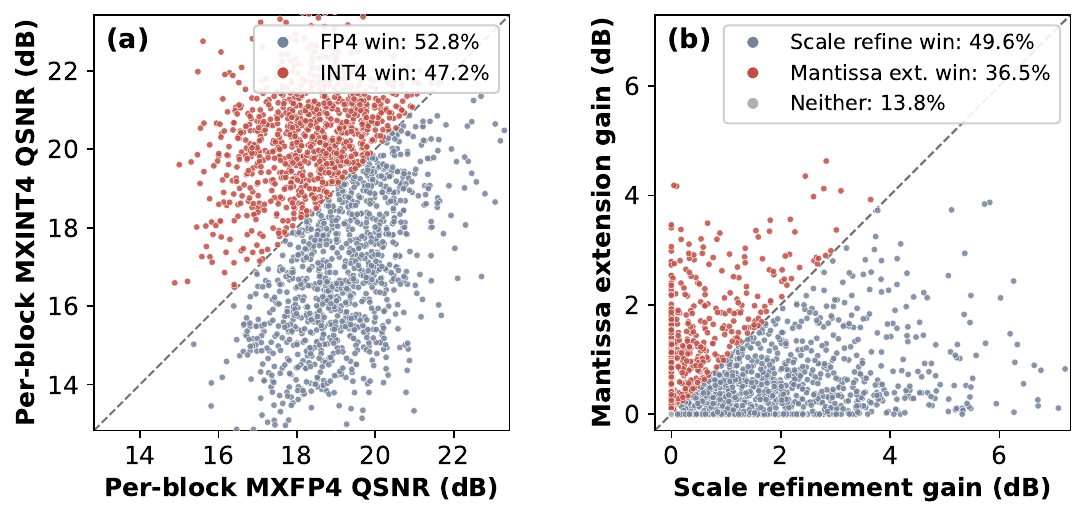}
  \caption{Block-level quantization heterogeneity in Llama-3.1-8B weights (block size 32).
    (a)~Format: \MXFPfour vs.\ \MXINTfour QSNR.
    (b)~Enhancement: scale refinement vs.\ block-maximum extension.
    Points above the diagonal favor \INTfour and mantissa extension, respectively.}
  \label{fig:block_diversity}
\end{figure}

\textbf{Weight-side block-level heterogeneity.}\label{sec:block_diversity}
\cref{fig:block_diversity} reveals two orthogonal forms of block-level heterogeneity in weights.
We quantize 2000 sampled blocks of Llama-3.1-8B and evaluate each quantized block using quantization signal-to-noise ratio (QSNR).
Panel (a) plots \MXFPfour against \MXINTfour, panel (b) the QSNR gain each enhancement adds, and each block is assigned to the option with the higher QSNR.
\emph{Format heterogeneity} sets whether \FPfour or \INTfour better fits a block: at the tensor level \MXFPfour beats \MXINTfour on most tensors~\cite{chen2025intvsfp}, but this tensor-level average does not reflect the per-block distribution, where 52.8\% of blocks favor \FPfour and 47.2\% favor \INTfour.
\emph{Enhancement heterogeneity} sets whether a block gains more from scale refinement or block-maximum extension: 49.6\% of blocks prefer the former, 36.5\% the latter, and the remaining 13.8\% gain little from either.
The two axes are orthogonal, and the diversity holds across all four models and at block size 16, so no single static assignment is optimal across all blocks.
Because weights are static, \AdaMX can search each block offline and route both axes per block (\cref{sec:weight_format}).

\textbf{Activation-side online encoding constraints.}
Activations are produced online each forward pass, making the offline per-block search unsuitable for activation encoding, and they consistently favor \FPfour over \INTfour at the tensor level~\cite{chen2025intvsfp,smoothquant}.
\AdaMX therefore fixes the format and instead refines the scale with a single deterministic rounding (\cref{sec:s1_encoding}), then extends the block maximum online.
Online block-maximum promotion requires additional metadata: storing a per-block index to locate the element costs 5~bits~\cite{mxplus}, which exceeds the available metadata budget, while appending mantissa bits to the stored \FPfour value~\cite{m2xfp} succeeds only when the higher-precision result shares the same leading bits.
When it does not, prior work clamps to a neighborhood of the stored value and loses precision~\cite{m2xfp}.
\AdaMX instead encodes the block maximum losslessly (\cref{sec:n1_encoding}).

\begin{figure}[t]
  \centering
  \includegraphics[width=\linewidth]{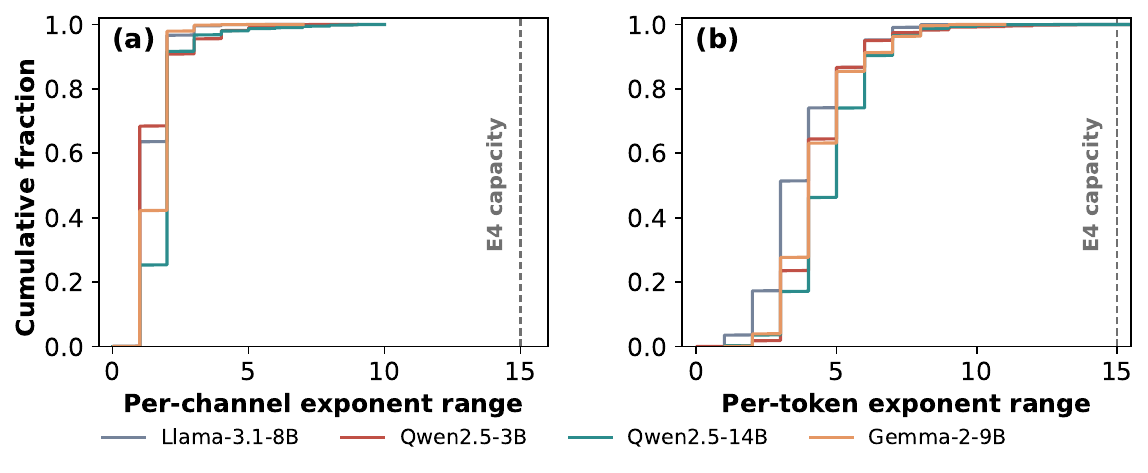}
  \caption{CDF of exponent range across four models (block size 32); the dashed line marks the \Efour capacity of 15.}
  \label{fig:exp_overprov}
\end{figure}

\textbf{Metadata capacity for adaptation.}\label{sec:exp_overprov}
This adaptation preserves the original per-block storage budget because the MX exponent is overprovisioned.
\cref{fig:exp_overprov} plots the per-block exponent range.
For weights, over 90\% of output channels span a range of 3 or less and none exceeds 11, well inside \Efour's capacity of 15, once a per-channel bias absorbs the inter-channel offset~\cite{smoothquant,awq} at a cost of only 0.002~bits/element.
For activations, a per-token bias computed online during WikiText-2 inference~\cite{wikitext} keeps more than 99.99\% of blocks within range on every model, with the clamped tail reaching 0.0086\% on Qwen2.5-3B and at most 0.0002\% on the others.
A 4-bit exponent thus suffices for both operands, leaving four spare bits per block that \AdaMX uses as metadata at no change in \EBW (\cref{sec:twolevel}).

\section{\AdaMX Design}
\label{sec:design}

\Cref{sec:motivation} established that the 8-bit MX scale field is overprovisioned for LLM workloads.
This section presents \AdaMX, a format family that reclaims these spare bits as per-block metadata.
We first introduce the two-level scale structure that enables narrower per-block exponents while preserving full dynamic range (\cref{sec:twolevel}), 
then detail the weight-side format with its 2-bit type routing across four format-enhancement combinations (\cref{sec:weight_format}), 
followed by the activation-side format, which combines a 1-bit scale refinement with a lossless \FPsix block-max extension (\cref{sec:act_format}).
The \AdaMX family provides two operating points that share identical encoding and hardware but differ in block size: \AdaMXSixteen at 4.5~bits/element (block size 16) and \AdaMXThirtyTwo at 4.25~bits/element (block size 32).

\subsection{Two-Level Scale with Anchored Bias}
\label{sec:twolevel}

Both \AdaMX formats replace the 8-bit \Eeight shared exponent with a narrower per-block exponent anchored by a coarser bias.
The effective scale for each block is
\begin{equation}
  S = 2^{\,b_{\mathrm{bias}} \;+\; e_{\mathrm{block}}},
  \label{eq:twolevel}
\end{equation}
where $b_{\mathrm{bias}}$ is an 8-bit integer set at a coarser granularity and $e_{\mathrm{block}}$ is the per-block residual exponent in \Efour. Activations refine this scale with a 1-bit mantissa.

\textbf{Weights.}
$b_{\mathrm{bias}}$ is set to the minimum block exponent within each output channel, making every residual $e_{\mathrm{block}}$ nonnegative;
the per-channel range stays inside \Efour's 0--15.

\textbf{Activations.}
$b_{\mathrm{bias}}$ is set per token, computed online by tracking the minimum of the per-block scale exponents as they are produced by the encoder, with $e_{\mathrm{block}}$ the unsigned residual.

\subsection{Weight Format: \WFmt}
\label{sec:weight_format}

\begin{figure}[t]
  \centering
  \includegraphics[width=\columnwidth]{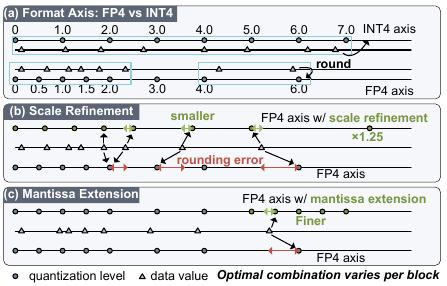}
  \caption{The two axes of \Ttwo routing. 
  (a)~\FPfour and \INTfour differ in grid structure; 
  the better format varies per block. 
  (b)~Scale refinement adjusts all grid points. 
  (c)~Mantissa extension refines only the block maximum.}
  \label{fig:t2_axes}
\end{figure}
\subsubsection{Bit Layout and \Ttwo Codespace}
\label{sec:t2_codespace}

The weight-side metadata byte (\cref{fig:figure1_data_formats}) packs three fields: \Efour (4~bits) for the residual exponent relative to the per-channel bias, \Mttwo (2~bits) as a metadata payload whose interpretation depends on \Ttwo, and \Ttwo (2~bits) for type routing each block to one of four \{format, enhancement\} combinations (\cref{tab:t2_codespace}).

\begin{table}[t]
\centering
\scriptsize
\renewcommand{\arraystretch}{1.05}
\caption{\Ttwo codespace: four format-enhancement combinations.}
\label{tab:t2_codespace}
\setlength{\tabcolsep}{6pt}
\begin{threeparttable}
\begin{tabular}{@{}c >{\raggedright\arraybackslash}p{0.36\linewidth} >{\raggedright\arraybackslash}p{0.42\linewidth}@{}}
\Xhline{1.2pt}
\Ttwo & Format + Enhancement & \Mttwo  \\
\hline
00 & \FPfour + block-max mantissa ext. & Extra \FPsix mantissa (2 bits) \\
01 & \FPfour + scale refinement & Ratio $\in\{1.0,\,1.25,\,1.5,\,1.75\}$ \\
10 & \INTfour + scale refinement & Same ratio set \\
11 & \mbox{\INTfour + block-max mantissa ext.} & Extra INT6 mantissa (2 bits)\tnote{1} \\
\Xhline{1.2pt}
\end{tabular}
\begin{tablenotes}\footnotesize
\item[1] INT4 in MX uses sign-magnitude representation. The 2-bit extension appends magnitude bits.
\end{tablenotes}
\end{threeparttable}
\end{table}

\textbf{Format axis.}
\Cref{sec:block_diversity} established that \FPfour and \INTfour each suit roughly half of all weight blocks within a tensor: the \FPfour grid favors blocks where most values cluster near zero with a few large outliers, while \INTfour's uniform grid favors flatter magnitude distributions. 
\Ttwo therefore routes the format choice per-block (\cref{fig:t2_axes}(a)), capturing diversity that tensor-level or layer-level assignment cannot.
The two grids allocate quantization levels differently across the representable range: \FPfour allocates more levels near zero and fewer levels at large magnitudes, whereas \INTfour distributes its levels uniformly.
Blocks concentrated near zero therefore tend to incur lower error under \FPfour, whereas blocks with flatter magnitude distributions tend to favor \INTfour.

\textbf{Enhancement axis.}
Two metadata enhancement techniques target different error sources (\cref{sec:bg:prior}).
Scale refinement adjusts the shared scale by a fractional ratio between adjacent power-of-two bins, with \Mttwo indexing four values $\{1.0, 1.25, 1.5, 1.75\}$.
Block-maximum mantissa extension instead promotes a single element. 
Because the shared scale anchors to the block maximum, that element lands in the format's top octave, where the grid is coarsest and its error dominates the block MSE.
\Mttwo extends its precision from 4-bit to 6-bit.
\Cref{fig:t2_axes}(b) applies scale refinement.
Every level moves, so each value in the block lands closer to a level and its rounding error shrinks.
\Cref{fig:t2_axes}(c) extends the mantissa instead.
Only the levels around the block maximum become finer, so that element's error shrinks and the rest of the grid is untouched.
\Cref{sec:block_diversity} showed that neither technique uniformly dominates at per-block granularity.
All prior MX metadata enhancement formats commit to a single enhancement axis at design time; \Ttwo decouples the format and enhancement axes, 
routing each block to the combination that reduces the most error.

\begin{algorithm}[t]
\caption{\WFmt weight block encoding.
$P$ is the largest power-of-two value in the element format;
$R$ is the ratio set for scale refinement;
$N_4(\hat{x}_j)$ is the 4-value decode neighborhood;
$\arg\max_i$ returns the lowest index on ties.}
\label{alg:weight_encoding}
\setstretch{1.05}
\footnotesize

\KwIn{FP16 weight block $\mathbf{w} \in \mathbb{R}^{k}$, per-channel bias $b_{\mathrm{bias}}$}
\KwOut{Metadata $[\Efour \mid \Mttwo \mid \Ttwo]$; quantized block $\hat{\mathbf{w}}$}

$\Rset \gets \{1,\,1.25,\,1.5,\,1.75\}$\;
$e_0 \gets \lfloor \log_2(\max_i |w_i| / P) \rfloor - b_{\mathrm{bias}}$\;
$\mathit{best} \gets +\infty$\;

\For{$b \in \{-1,\,0,\,+1\}$}{
  $S \gets 2^{\,b_{\mathrm{bias}} + e_0 + b}$\;
  $\mathbf{x} \gets \mathbf{w}/S$\;

  \For(\tcp*[f]{\Ttwo Type codeword}){$t \in \{00,\,01,\,10,\,11\}$}{
    $\tau \gets \FPfour$ if $t \in \{00,\,01\}$, else $\INTfour$\tcp*[r]{format}

    \uIf(\tcp*[f]{scale refinement}){$t \in \{01,\,10\}$}{
      $E(r) \gets \left\|\mathbf{x} - \Qround{\mathbf{x}/r}{\tau}\,r\right\|^2,\quad r \in \Rset$\;
      $r^* \gets \argmin_{r \in \Rset} E(r)$\;
      $m \gets \mathrm{index}(r^*)$\;
      $\hat{\mathbf{w}}_t \gets S \cdot \Qround{\mathbf{x}/r^*}{\tau}\,r^*$\;
    }
    \Else(\tcp*[f]{block-max mantissa ext.}){
      $\tau_6 \gets \FPsix$ if $t = 00$, else $\mathrm{INT6}$\;
      $\hat{\mathbf{x}} \gets \Qround{\mathbf{x}}{\tau}$\tcp*[r]{quant to 4-bit}
      $j \gets \arg\max_i |\hat{x}_i|$\tcp*[r]{block-max index}
      $q_j \gets \mathrm{clamp}(\Qround{x_j}{\tau_6}, \Nbr{\hat{x}_j})$\;
      $m \gets \mathrm{index}(q_j,\Nbr{\hat{x}_j})$\tcp*[r]{2-bit neighborhood index}
      $\hat{x}_j \gets q_j$\;
      $\hat{\mathbf{w}}_t \gets S \cdot \hat{\mathbf{x}}$\;
    }

    $\mathit{mse} \gets k^{-1}\|\mathbf{w} - \hat{\mathbf{w}}_t\|^2$\;

    \If{$\mathit{mse} < \mathit{best}$}{
      $(\mathit{best}, \Efour, \Mttwo, \Ttwo, \hat{\mathbf{w}}^*)
      \gets (\mathit{mse}, e_0+b, m, t, \hat{\mathbf{w}}_t)$\;
    }
  }
}

\KwRet $[\Efour \mid \Mttwo \mid \Ttwo]$, $\hat{\mathbf{w}}^*$\;
\end{algorithm}

\subsubsection{Encoding}
\label{sec:weight_encoding}

\Cref{alg:weight_encoding} encodes each weight block into the byte-aligned $[\Efour\,|\,\Mttwo\,|\,\Ttwo]$ metadata.
For each block of $k$ elements, the encoder evaluates all four \Ttwo candidates at three exponent offsets $b \in \{-1, 0, +1\}$, giving twelve candidates per block; the lowest per-block MSE selects the stored \Efour, \Mttwo, and \Ttwo.
Because weights are static, this exhaustive selection runs once offline.
The exponent-offset search (line 4) subsumes the choice between floor and ceil scale rounding~\cite{yang2025arith,m2xfp} 
with the winning offset folded into \Efour at no additional storage cost.

For scale refinement (lines 9--13), the encoder selects the ratio that minimizes per-block MSE under the effective scale $r \cdot S$; \Mttwo records its index. 
The ratio applies uniformly to all elements, so no per-element encoding is needed.

For mantissa-extension (lines 14--21), \Mttwo stores a 2-bit index that promotes the block-maximum element from 4-bit to 6-bit precision (\FPsix or INT6).
The encoder first quantizes all $k$ elements to 4-bit and locates the block maximum in this 4-bit space so the decoder can recover the same element index (ties broken by taking the lowest index). 
To preserve alignment between the stored 4-bit value and the refined 6-bit value, the 6-bit result is constrained to a 4-element neighborhood centered on the original 4-bit value's position in the 6-bit grid, with \Mttwo selecting among them. 
For example, if the block maximum quantizes to \FPfour value $6.0$, the four \FPsix candidates are $\{5.5, 6.0, 6.5, 7.0\}$. 
This neighborhood does not cover every round-to-nearest 6-bit target. The weight encoder mitigates this through exhaustive selection (\cref{alg:weight_encoding}): if a block's optimal 6-bit target falls outside the neighborhood, the encoder routes it to scale refinement instead. Activations cannot perform this exhaustive search online and therefore require an additional encoding mechanism (\cref{sec:n1_encoding}).

\begin{figure}[t]
  \centering
  \includegraphics[width=\columnwidth]{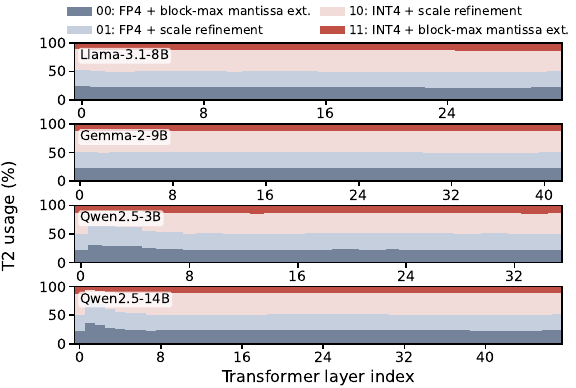}
  \caption{Per-layer \Ttwo codeword distribution for \AdaMX across four LLMs (block size 32).
   All four codewords are populated at every layer in every model.}
  \label{fig:per_layer_t2}
\end{figure}

\subsubsection{Decoding and Routing Behavior}
\label{sec:weight_decoding}
\Ttwo controls the decoding path.
For \Ttwo $\in \{01, 10\}$ (scale refinement), \Mttwo selects one of four ratio values from $\Rset$, and all $k$ block elements decode under the adjusted block scale.
For \Ttwo $\in \{00, 11\}$ (block-max mantissa extension), the block-maximum element is reconstructed at 6-bit precision from its 4-bit value and the 2-bit \Mttwo index; the remaining $k{-}1$ elements decode at 4-bit precision.
The decoder identifies this element by argmax over the decoded 4-bit magnitudes, the same rule the encoder used, with ties broken by the lowest index.
No position index is stored, so \Mttwo carries only the mantissa extension.

Per-block exhaustive selection on Llama-3.1-8B weights (block size 32) populates all four \Ttwo values: 18.6\% / 30.3\% / 35.3\% / 15.9\% for codewords 00, 01, 10, 11.
The roughly even FP/INT split (48.9\% / 51.1\%) is consistent with the format diversity of \cref{sec:block_diversity}, while the simultaneous use of both block-max mantissa extension (34.4\%) and scale refinement (65.6\%) enhancements reflects an axis that format selection alone cannot capture.
\Cref{fig:per_layer_t2} shows this distribution generalizes across four LLMs.
Per-block routing captures this diversity within the same 2-bit selector.
All four codewords are used with non-negligible frequency, indicating that no single static format-enhancement combination captures the observed block-level diversity.
Because \Ttwo spans exactly four codewords, the design therefore uses the complete 2-bit selector space without additional storage.

\subsection{Activation Format: \AFmt}
\label{sec:act_format}

\subsubsection{Bit Layout and Design Rationale}
\label{sec:act_layout}

The activation-side metadata byte consists of four fields (\cref{fig:figure1_data_formats}): \Efour (4~bits), \Mone (1~bit), \Mttwo (2~bits), and \None (1~bit).
\Efour and \Mone form the \AScale block scale: \Efour is the residual exponent relative to the per-token bias, and \Mone is a 1-bit scale mantissa (\cref{sec:s1_encoding}).
\Mttwo extends the block-maximum element from \FPfour to \FPsix~(E2M3) precision.
\None selects between two encoding neighborhoods that together cover every \FPsix round-to-nearest target (\cref{sec:n1_encoding}).

The activation side omits the \Ttwo type selector for two reasons: activation outliers favor \FPfour over \INTfour~\cite{chen2025intvsfp}, and online quantization rules out the costly per-block scale search.
\Mone thus refines the scale with a single rounding (\cref{sec:s1_encoding}), and the freed \Ttwo bits become \Mone and \None.

\subsubsection{Scale Refinement and Rounding}
\label{sec:s1_encoding}

\begin{figure}[t]
  \centering
  \includegraphics[width=\columnwidth]{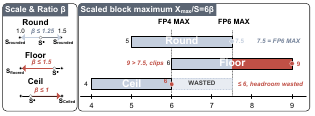}
  \caption{Scale-rounding effects: round-to-nearest fits the block maximum within \FPsix, floor clips, and ceil wastes headroom.}
  \label{fig:round_codesign}
\end{figure}

The \Mone mantissa turns the activation scale into \AScale, $S = 2^{\,b_{\mathrm{bias}} + e_{\mathrm{block}}}\!\cdot m$ with $m \in \{1, 1.5\}$, chosen to map the block maximum near \FPfour's max-norm $v_{\max} = 6$.

$S$ requires a rounding policy.
The original MX format rounds the shared scale down (floor)~\cite{rouhani2023mx,ocpmx}; later works use ceiling rounding to lower clipping~\cite{mishra2025mxfp8,yang2025arith} or adaptive block scaling~\cite{fouroversix}.
\AdaMX rounds to nearest, co-designed with the \FPsix block-max extension.
The ideal scale $S^{*}$ would land the block maximum exactly on $v_{\max} = 6$, but \Mone offers only $\{1, 1.5\}$ per octave, so the encoder must round $S^{*}$ to a nearby grid scale $S$.
Writing $\beta = S^{*}/S$ for the ratio of the ideal scale to the chosen scale, the block maximum lands at $6\beta$ instead (\cref{fig:round_codesign}): under-scaling ($\beta > 1$) overshoots $6$, and over-scaling ($\beta < 1$) falls short.
Round's worst ratio is the midpoint $1.25$ of the $\{1,1.5\}$ grid, so the block maximum reaches at most $6 \times 1.25 = 7.5$, exactly \FPsix's max-norm, which \None recovers with no clipping (\cref{sec:n1_encoding}).
Floor's worst ratio is $1.5$, reaching $6 \times 1.5 = 9$, which clips even in \FPsix; ceil holds the maximum below $6$ and leaves the \FPsix headroom unused.
Only round-to-nearest bounds the worst-case scaled block maximum at the maximum representable \FPsix value, demonstrating that the scale grid and block-maximum extension are co-designed.
With \AdaMXSixteen otherwise fixed, round gives the lowest perplexity: $6.86$, versus $6.90$ for ceil and $7.08$ for floor on Llama-3.1-8B, and $8.80$ versus $8.90$ and $9.07$ on Qwen2.5-3B.

Scale computation requires no division.
Write the block maximum as $\max_i|x_i| = m_a\,2^{e_a}$ with $m_a \in [1,2)$.
\FPfour's largest representable value is $v_{\max} = 6 = 1.5\cdot 2^2$.
The ideal scale is then $S^{*} = \max_i|x_i|/v_{\max} = (m_a/1.5)\,2^{\,e_a-2}$.
The factor $2^{\,e_a-2}$ is an exact power of two and only sets the exponent.
Rounding the ideal scale onto the $\{1,1.5\}\times 2^{\mathbb{Z}}$ grid thus reduces to rounding the mantissa factor $(m_a/1.5)$ onto the same grid.
Since $m_a \in [1,2)$, this factor spans $[2/3,\,4/3)$.
This range rounds to one of three grid points: $0.75$, $1$, and $1.5$.
Here $1$ and $1.5$ are the two grid values in one octave, and $0.75 = 1.5\cdot 2^{-1}$ is the next value one octave below.
Round-to-nearest splits at their midpoints $0.875$ and $1.25$.
Multiplying the two midpoints by $1.5$ gives the block-max thresholds $1.3125$ and $1.875$.
These thresholds split $m_a$ into three cases:

{\small
\renewcommand{\arraystretch}{0.9}%
\begin{equation}
(E, m) =
\begin{cases}
(e_a - 3,\ 1.5), & m_a < 1.3125,\\
(e_a - 2,\ 1.0), & 1.3125 \le m_a < 1.875,\\
(e_a - 2,\ 1.5), & m_a \ge 1.875,
\end{cases}
\label{eq:s1_round}
\end{equation}
}
where $E$ is the scale exponent, $m$ the \Mone\ mantissa, and $e_{\mathrm{block}} = E - b_{\mathrm{bias}}$ the stored residual.
The three cases round $(m_a/1.5)$ to $0.75$, $1$, and $1.5$ in turn.
Rounding to $0.75 = 1.5\cdot 2^{-1}$ lowers the exponent by one and yields the $e_a - 3$ row.
The encoder reads $e_a$ and compares the leading bits of $m_a$ against these two constants; the quantization unit realizes this with two comparators and a mux (\cref{sec:hw_aqu}).

\subsubsection{\None: Lossless \FPsix Encoding via Neighborhood Selection}
\label{sec:n1_encoding}

\Cref{sec:weight_format} introduced a constrained \FPsix encoding that concatenates the stored \FPfour bits with \Mttwo to address a 4-value neighborhood on the \FPsix grid. 
Writing $\mathrm{FP4\_idx} \in [0,7]$ and $\mathrm{FP6\_idx} \in [0,31]$ for the magnitude indices on the two grids (signs decode independently), the reconstruction is
\[
\mathrm{FP6\_idx} = \mathrm{FP4\_idx} \times 4 + \Mttwo + s,
\]
where the integer shift $s$ positions the 4-value window on the \FPsix grid. 
The constrained \FPsix encoding sets $s = -1$, the $[-1, +2]$ neighborhood used by previous work~\cite{m2xfp}.
The encoding loses precision when the unconstrained \FPsix round-to-nearest result falls outside this neighborhood.

To characterize the offset distribution, let $\Delta = \mathrm{FP6\_idx}_{\mathrm{u}} - \mathrm{FP4\_idx} \times 4$ be the offset of the unconstrained \FPsix result from the stored \FPfour index.
Inside \FPfour's range this offset stays in $\{-2, \ldots, +2\}$; round scaling (\cref{sec:s1_encoding}) then drives the block maximum to $7.5$, three \FPsix steps above \FPfour's ceiling of $6.0$, adding a sixth offset $+3$.
These six offsets cannot all be covered by the 4-value constrained window, so two of them require clamping and lose precision.
In value terms, the block maximum rounds to one of $\{5.0, 5.5, 6.0, 6.5, 7.0, 7.5\}$ on the \FPsix\ grid, and the constrained window $\{5.5, 6.0, 6.5, 7.0\}$ omits $5.0$ and $7.5$.
Increasing the grid precision alone is insufficient: FP7 produces eleven possible offsets, exceeding the coverage of even a 3-bit neighborhood selector.

\begin{figure}[t]
  \centering
  \includegraphics[width=\columnwidth]{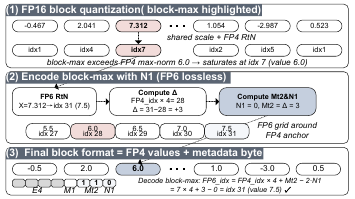}
  \caption{\AFmt\ encoding flow.}
  \label{fig:n1_flow}
\end{figure}

To cover all six offsets, \None selects between two overlapping 4-value neighborhoods, $\{0, +1, +2, +3\}$ for $\None = 0$ and $\{-2, -1, 0, +1\}$ for $\None = 1$, whose union spans the complete offset set.
\Mttwo indexes each neighborhood in order, and the two share offsets $0$ and $+1$ so every \FPsix round-to-nearest result is reachable (\cref{fig:n1_flow}).
The encoder picks $\None = 0$ and $\Mttwo = \Delta$ when $\Delta \geq 0$; otherwise $\None = 1$ and $\Mttwo = \Delta + 2$.
The decoder uses one shift, one add, and one conditional subtract:
\[
\mathrm{FP6\_idx} = \mathrm{FP4\_idx} \times 4 + \Mttwo - 2 \cdot \None.
\]
Relative to constrained \FPsix~\cite{m2xfp} decoding, this design adds only one fixed-offset subtraction.
Six offsets map to six $(\Mttwo, \None)$ codes, so the encoding is lossless.

This lossless encoding allows \None to eliminate block-maximum clamping, while the prior constrained \FPsix clamps $13\%$ and a finer FP7 grid clamps $11\%$.
Eliminating these clamped cases reduces the block-maximum squared error by $2.2\times$, from $0.047$ to $0.021$, independent of model.
With only the block-max encoding varied under \AScale round scaling, \None also achieves the lowest perplexity on every model, ahead of both the prior constrained \FPsix and the FP7 alternative:

\begin{center}
\scriptsize
\setlength{\tabcolsep}{4pt}
\makebox[\linewidth]{%
\begin{tabular}{@{}lcc@{}}
\hline
Encoding & MSE & Clamp \\
\hline
constr.\ \FPsix & 0.047 & 13\% \\
\FPsix{+}\None & \textbf{0.021} & \textbf{0\%} \\
\hline
\end{tabular}%
\hfill
\begin{tabular}{@{}lccc@{}}
\hline
Model & constr.\ \FPsix & FP7 & \FPsix{+}\None \\
\hline
Qwen2.5-3B & 8.93 & 8.83 & \textbf{8.80} \\
Llama-3.1-8B & 6.94 & 6.88 & \textbf{6.86} \\
Qwen2.5-14B & 5.91 & 5.86 & \textbf{5.84} \\
\hline
\end{tabular}%
}
\end{center}

FP7 also requires more hardware: it needs a wider mantissa multiplier and data-dependent clamp logic, whereas \None keeps the coarser \FPsix datapath and decodes with one fixed-offset subtract.
Thus, the lossless FP6 encoding is both more accurate and more hardware-efficient than a finer FP7 grid that still requires clamping.

\subsection{Choosing the Block Size}
\label{sec:blocksize}

\begin{figure}[t]
  \centering
  \includegraphics[width=\columnwidth]{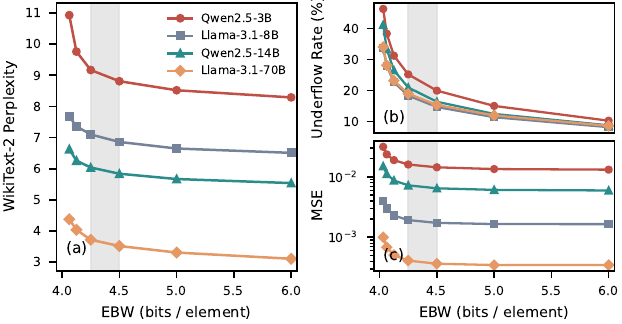}
  \caption{(a) WikiText-2 perplexity, (b) FP4 underflow rate, and (c) reconstruction MSE vs \EBW\ on four LLMs; the shaded band marks block sizes 16 and 32.}
  \label{fig:choose_block_size}
\end{figure}

\AdaMX's hardware supports two block sizes, 16 and 32, on one datapath.
We justify both with a block-size sweep from 4 to 256 on four LLMs, plotting perplexity, the FP4 underflow rate, and the block reconstruction MSE against the resulting \EBW\ ($=4+8/N$) (\cref{fig:choose_block_size}).

All three metrics degrade as the block size increases.
A larger block shares one scale across a wider range, so more of its small elements fall below FP4's smallest level and quantize to zero.
The underflow rate is the fraction of nonzero elements lost this way, and it reaches 35 to 46\% at the largest block.
Further reductions below block size 16 provide only marginal accuracy improvements while increasing \EBW\ by 0.5 to 1.5 bits per element for each halving.

Block sizes 16 and 32 lie near the knee of all three curves.
Neither dominates: \AdaMXSixteen\ (4.5 \EBW) is more accurate, \AdaMXThirtyTwo\ (4.25 \EBW) uses less memory.
Supporting both on one datapath makes the block size a flexible operating point, not a fixed design parameter.

\section{Architecture}
\label{sec:architecture}

This section describes the \AdaMX\ hardware,
which integrates AdaMX into an output-stationary systolic array (SA) with lightweight modifications.
A fixed 32-lane MAC microarchitecture supports runtime-selectable block sizes $K \in \{16, 32\}$,
so one design serves both operating points.
\cref{sec:eval} reports the corresponding power, performance, and area results.

\subsection{Hardware Overview}
\label{sec:hw_overview}
 
\begin{figure}[t]
  \centering
  \includegraphics[width=\columnwidth]{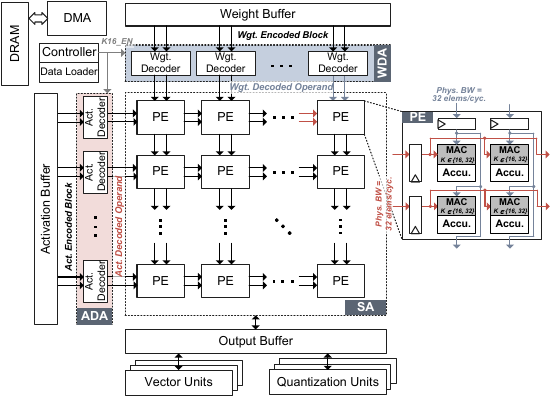}
  \caption{\AdaMX\ architecture overview.
  Each fixed 32-lane dot-product MAC supports both block sizes.}
  \label{fig:hw_overview}
\end{figure}

\cref{fig:hw_overview} presents the \AdaMX\ architecture, which uses an $M{\times}N$ output-stationary systolic array (SA).
Each \textbf{PE} computes a $2{\times}2$ output tile with four fixed 32-lane dot-product \textbf{MAC blocks}, each with a private FP32 accumulator.
The four MACs share two activation-row and two weight-column operand registers, halving the operand-register count per MAC relative to four single-MAC tiles.

At the SA boundary, the \textbf{activation decoder array (ADA)} places one decoder per row and the \textbf{weight decoder array (WDA)} one per column.
Each decoder consumes one encoded operand per cycle and broadcasts 32 decoded operands along its row or column (\cref{sec:hw_decoder}).
For an $M{\times}N$ array this placement reduces decoder replication from $MN$ instances to $M{+}N$.
Each MAC pairs a 32-lane base path, built from asymmetric 2-by-3-bit element multipliers, with a per-block correction sidepath for block-maximum mantissa extension, followed by the scale-refinement and normalization stages (\cref{sec:hw_mac}).
The output buffer feeds the vector units for the non-linear operations and the \textbf{activation quantization unit}, which re-encodes the layer's FP32 outputs into \AFmt\ blocks for the next layer (\cref{sec:hw_aqu}).
The controller manages the dataflow, the DMA engine handles off-chip memory access, and the data loader drives operands from the buffers into the SA.
A global \texttt{K16\_EN} signal selects the block size for the whole pass:
$K{=}32$ maps one 32-element block onto the physical lanes,
while $K{=}16$ packs two independent 16-element blocks on the same lanes (\#0 on 0--15, \#1 on 16--31) at equal throughput.

\subsection{Decoder Lane}
\label{sec:hw_decoder}

\begin{figure}[t]
  \centering
  \includegraphics[width=\columnwidth]{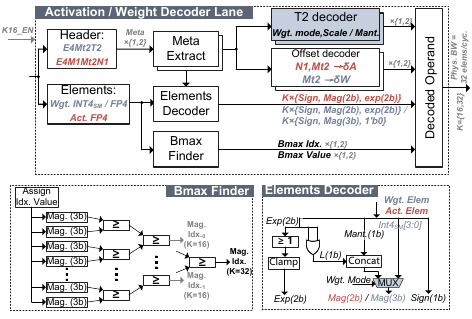}
  \caption{\AdaMX\ decoder,
  supporting both $K{=}16$ and $K{=}32$.}
  \label{fig:hw_decoder}
\end{figure}

\cref{fig:hw_decoder} details one decoder lane.
Each lane consumes one encoded operand per cycle
and produces two outputs:
32 decoded operands for the base path,
and a per-block control bundle that drives the MAC sidepaths.
The activation and weight lanes share most logic;
\cref{fig:hw_decoder} shows the combined structure,
with weight-specific blocks shaded in blue and activation-specific signals in red.
At $K{=}16$ the operand carries two headers.
 
\textbf{Element decode.}
Each element decoder converts a 4-bit element into a sign,
a clamped exponent,
and a mantissa.
For \FPfour\ elements,
an explicit leading bit folds the subnormal and smallest-normal values into a single mantissa-and-shift form,
so one datapath covers the full grid without a subnormal branch.
For \FPfour, each lane emits a sign, a 2-bit clamped exponent, and a 2-bit significand formed by the explicit leading bit and the stored mantissa bit.
In \INTfour\ weight mode the shared element interface instead carries a 3-bit sign-magnitude value with the exponent field tied to zero, selected by Wgt.Mode.
At $K{=}16$ each 16-block applies its own Wgt.Mode to its own 16 lanes.

\textbf{T2 decode.}
The weight header is decoded once per block to generate the MAC control signals.
Wgt.Mode $= \Ttwo[1]$ selects the \FPfour\ or \INTfour\ base path.
The ratio modes ($\Ttwo[0] \oplus \Ttwo[1]$) route $\mathrm{Mt2}_w$
to the scale-refinement field $\mathrm{Mt2}_{\mathrm{eff}}$;
the block-max modes instead direct $\mathrm{Mt2}_w$
to the offset decoder
and force $\mathrm{Mt2}_{\mathrm{eff}}$ to zero.
The activation header carries no \Ttwo.
Its \Mone\ bit drives the second scale-refinement stage (\cref{sec:hw_mac}), while $\mathrm{Mt2}_a$ with \None\ feed the offset decoder and the block exponent is forwarded to the normalization stage.

\textbf{Block-max finder and offset decode.}
Both lanes locate the block-max element
by reducing the stored magnitudes in a comparator tree,
breaking ties toward the lowest index
so that the decoder and encoder select the same element.
The tree is shared across block sizes.
At $K{=}32$ its root gives the single block maximum.
At $K{=}16$ its penultimate stage already holds the winners of lanes 0--15 and lanes 16--31,
so one tree returns both block maxima without a second instance.
The exponent, sign, and index at each block-max position
are read from the local decoded arrays and placed in the bundle.
A small offset decoder reconstructs the signed mantissa offset $\delta_W$ or $\delta_A$
from the enhancement field,
$\mathrm{Mt2}_w$ on the weight side (\cref{sec:weight_encoding})
and $\mathrm{Mt2}_a$ with \None\ on the activation side (\cref{sec:n1_encoding}).

\subsection{MAC Microarchitecture}
\label{sec:hw_mac}
 
\begin{figure}[t]
  \centering
  \includegraphics[width=\columnwidth]{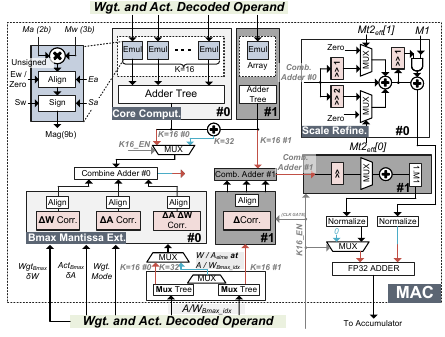}
  \caption{\AdaMX\ MAC microarchitecture. Light-gray and dark-gray mark the two per-block back-ends.}
  \label{fig:hw_mac}
\end{figure}

\cref{fig:hw_mac} shows the \AdaMX\ MAC.
The MAC receives two 32-element decoded operand vectors, one from the ADA and one from the WDA, together with the per-block control signals derived from their metadata.
The base path contains 32 parallel 2-by-3-bit unsigned element multipliers in the \textbf{element multiplier array}.
After exponent alignment and sign application, a 32-input signed adder tree reduces the products to one signed fixed-point partial sum.
At $K{=}16$ the tree's penultimate stage already yields the two 16-block sums, so one tree serves both block sizes.
The sidepath produces up to three per-block scalar correction terms in the \textbf{block-maximum mantissa extension} stage,
which are aligned and summed with the base-path partial sum in the \textbf{combine adder}.
Two cascaded shift-add stages in \textbf{scale refinement} then apply the weight and activation scale factors, and \textbf{normalize} stage produces the FP32 output.
The \Ttwo\ and \None\ metadata fields are decoded once per block in the WDA and ADA respectively;
the MAC receives only the derived control signals,
the mantissa offsets $\delta_W$ and $\delta_A$,
and the block-max indices $A_{\text{Bmax\_idx}}$ and $W_{\text{Bmax\_idx}}$ that drive the W-Pick and A-Pick blocks.

\textbf{Block-maximum mantissa extension.}
Both weight and activation blocks may carry extra mantissa bits on their block-max element,
encoded as offsets $\delta_W$ and $\delta_A$ that augment the base values.
This yields three correction terms beyond the base-path accumulation:
\begin{equation}
\small
\label{eq:bilinear}
\begin{aligned}
\sum_i W_i A_i
= \sum_i W_{\mathrm{base},i} A_{\mathrm{base},i}
  &+ W_{\mathrm{base},a_{\mathrm{bmax}}} \delta_a
   + A_{\mathrm{base},w_{\mathrm{bmax}}} \delta_w \\
  &+ \delta_w \delta_a \,
  \mathbb{I}\!\left\{w_{\mathrm{bmax}} = a_{\mathrm{bmax}}\right\}.
\end{aligned}
\end{equation}


The three corrections are realized in $\Delta A$ Corr,
$\Delta W$ Corr,
and $\Delta A \Delta W$ Corr (\cref{fig:hw_mac}),
each a lightweight scalar MAC mirroring one base-path lane.
$\Delta A$ Corr and $\Delta W$ Corr take a base operand from the opposite side at the block-max index,
supplied by the W-Pick and A-Pick blocks ($W_{\text{elem}}$ at $A_{\text{Bmax\_idx}}$ and $A_{\text{elem}}$ at $W_{\text{Bmax\_idx}}$ respectively);
$\Delta A \Delta W$ Corr multiplies the two scalar offsets directly.
The three correction enables are gated by \Ttwo:
$\Delta A$ Corr is always enabled,
$\Delta W$ Corr is enabled for $\Ttwo \in \{00, 11\}$,
and $\Delta A \Delta W$ Corr is enabled only when $\Delta W$ Corr is active
and the two block-max indices coincide.
 
\textbf{Scale refinement.}
Both operands carry a per-block scale factor, and two cascaded shift-add stages apply them without a multiplier.
For weight blocks in ratio mode ($\Ttwo \in \{01, 10\}$),
the first stage applies the ratio factor $\rho = 1 + \mathrm{Mt2}_w / 4 \in \{1, 1.25, 1.5, 1.75\}$ for $\mathrm{Mt2}_w \in \{00, 01, 10, 11\}$
to the combine-adder output $s$:
\begin{equation}
\label{eq:ratio}
u = \rho  s = s + \mathrm{Mt2}_{\text{eff}}[1] \cdot (s \gg 1) + \mathrm{Mt2}_{\text{eff}}[0] \cdot (s \gg 2).
\end{equation}
The base term $s$ is added unconditionally;
$\mathrm{Mt2}_{\text{eff}}[1]$ and $\mathrm{Mt2}_{\text{eff}}[0]$ gate the $\times 0.5$ and $\times 0.25$ contributions via 1-bit and 2-bit right shifts.
In block-max mantissa extension modes ($\Ttwo \in \{00, 11\}$),
$\mathrm{Mt2}_{\text{eff}}$ is forced to zero and the stage passes $s$ through unchanged.
Every activation block carries its own factor $\mu = 1 + \Mone / 2 \in \{1, 1.5\}$, the \AScale\ mantissa (\cref{sec:s1_encoding}).
The second stage applies it with one more gated shift-add, $v = \mu u = u + \Mone \cdot (u \gg 1)$, so the MAC realizes the product $\rho \mu$ of the two factors.

\textbf{Normalization.}
The Normalization stage converts the fixed-point result $v$ to FP32 and incorporates the per-block scale.
On each side, the two-level decoding (\cref{sec:twolevel}) gives a sideband bias $b_{\mathrm{bias}}$ and per-block exponent $e_{\mathrm{block}}$.
After integer-to-float conversion,
a single integer add into the FP32 biased-exponent field accumulates the four contributions $b_{\mathrm{bias},w} + e_{\mathrm{block},w} + b_{\mathrm{bias},a} + e_{\mathrm{block},a}$,
applying the full per-block scale without any floating-point multiplication.
The resulting FP32 value is summed into the MAC's FP32 accumulator.
 
\textbf{Runtime block-size support.}
In \cref{fig:hw_mac} the light-gray block \#0 and dark-gray block \#1 are the two per-block back-ends,
selected by the \texttt{K16\_EN} muxes.
The base path is shared and always processes all 32 lanes:
the adder tree and the W-Pick and A-Pick mux trees each expose the two 16-block results at their penultimate stage,
with the root used for $K{=}32$.
At $K{=}16$ both back-ends run, one per 16-block, keeping all 32 lanes at full throughput;
at $K{=}32$ the \#1 back-end is held at zero.

\subsection{Activation Quantization Unit}
\label{sec:hw_aqu}

\begin{figure}[t]
  \centering
  \includegraphics[width=\columnwidth]{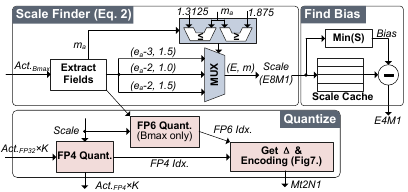}
  \caption{\AdaMX\ activation quantization unit.
  Blue and red mark the hardware-friendly Scale Finder and FP4/FP6 Quant.}
  \label{fig:hw_aqu}
\end{figure}
 
The activation quantization unit re-encodes the FP32 outputs into \AFmt\ blocks for the next layer (\cref{fig:hw_aqu}).
A running maximum over the streaming FP32 outputs determines the block scale.
The \textbf{Scale Finder} reads the amax leading bits (\cref{eq:s1_round}) with two comparators and a mux, giving the \AScale\ exponent $E$ and \Mone\ mantissa.
\textbf{FP4 Quant} maps every element to \FPfour\ by round-to-nearest.
The block-maximum index then comes from the quantized \FPfour\ magnitudes, with ties resolved toward the lowest index so the decoder locates the same element.
\textbf{FP6 Quant} re-quantizes that element, encoding its offset as $\Delta \to (\Mttwo, \None)$ (\cref{sec:n1_encoding}).
Each block's absolute 8-bit exponent $E$ is buffered in the scale cache while a running $\min$ over the token tracks the per-token bias $b_{\mathrm{bias}}$.
Once the token completes, the cache drains, and each stored $E$ is converted during cache draining to the 4-bit residual $e_{\mathrm{block}} = E - b_{\mathrm{bias}}$ (\cref{sec:twolevel}) as its final \AFmt\ block is written.
The pipeline is streaming and overlaps with SA computation without stalling the array.

\textbf{Division-free quantization.}
Quantization conceptually scales each element by $1/S = 1/(2^E\!\cdot m)$ and rounds it to the \FPfour\ grid $\{0, 0.5, 1, 1.5, 2, 3, 4, 6\}$; the hardware implements this without a divider or multiplier.
The $2^E$ factor is removed by subtracting $E$ from the element exponent.
Rounding then reduces to comparisons against the grid midpoints $\{0.25, 0.75, 1.25, 1.75, 2.5, 3.5, 5.0\}$, the round-to-nearest boundaries.
The \Mone\ factor $m \in \{1, 1.5\}$ scales the grid, so \Mone\ selects one of two constant boundary sets through a mux.
For example, a magnitude $5.0$ with $E = 0$ and $m = 1.5$ has exact value $5.0/1.5 = 3.33$, which rounds to $3.0$.
The hardware instead places $5.0$ between the scaled boundaries $1.5 \times 2.5 = 3.75$ and $1.5 \times 3.5 = 5.25$, producing the same result without division.
The block-max \FPsix\ promotion reuses this comparison.
Because \cref{eq:s1_round} lands the block maximum in $[5.0, 7.5)$, FP6 Quant resolves its six reachable codes with only five comparisons.
The \Mone\ scale refinement thus adds no arithmetic on the encode path.

\begin{table*}[t]
\centering
\scriptsize
\renewcommand{\arraystretch}{1.2}
\setlength{\tabcolsep}{2pt} 
\caption{Downstream accuracy (\%) on six zero-shot commonsense tasks and five-shot MMLU, for four models ordered small to large.
\textbf{Bold} marks the best low-bit result in each model;
the Recovery rows give the share of the accuracy gap from \MXFPfour\ to FP16 that each format recovers on the average rows.
}
\label{tab:downstream}
\resizebox{\textwidth}{!}{
\begin{tabular}{l *{4}{|cccccc}}
\Xhline{1.2pt}
& \multicolumn{6}{c|}{Qwen2.5-3B} & \multicolumn{6}{c|}{Llama-3.1-8B} & \multicolumn{6}{c|}{Qwen2.5-14B} & \multicolumn{6}{c}{Llama-3.1-70B} \\
\cline{2-7}\cline{8-13}\cline{14-19}\cline{20-25}
Task & FP16 & \AdaMXSixteen & \AdaMXThirtyTwo & \MMXFP & \NVFPfour & \MXFPfour
     & FP16 & \AdaMXSixteen & \AdaMXThirtyTwo & \MMXFP & \NVFPfour & \MXFPfour
     & FP16 & \AdaMXSixteen & \AdaMXThirtyTwo & \MMXFP & \NVFPfour & \MXFPfour
     & FP16 & \AdaMXSixteen & \AdaMXThirtyTwo & \MMXFP & \NVFPfour & \MXFPfour \\
\EBW & 16 & 4.5 & 4.25 & 4.5 & 4.5 & 4.25 & 16 & 4.5 & 4.25 & 4.5 & 4.5 & 4.25 & 16 & 4.5 & 4.25 & 4.5 & 4.5 & 4.25 & 16 & 4.5 & 4.25 & 4.5 & 4.5 & 4.25 \\
\hline
\multicolumn{25}{l}{\textbf{Commonsense (zero-shot)}}\\
\hline
ARC-C & 47.10 & \textbf{45.65} & 43.86 & 44.11 & 44.37 & 43.17 & 55.12 & 52.56 & 51.62 & 49.91 & \textbf{53.24} & 45.56 & 59.30 & \textbf{59.22} & 58.45 & 58.37 & 56.57 & 53.67 & 61.01 & \textbf{61.60} & 61.18 & 61.43 & 60.41 & 55.72 \\
HellaSwag & 73.51 & \textbf{71.44} & 70.81 & 70.12 & 70.37 & 66.44 & 79.30 & \textbf{78.12} & 77.78 & 77.36 & 77.86 & 74.19 & 82.94 & \textbf{81.61} & 81.38 & 81.35 & 81.37 & 79.40 & 85.71 & \textbf{85.06} & 84.96 & 83.80 & 84.96 & 81.44 \\
PIQA & 78.45 & \textbf{77.42} & 77.26 & 76.93 & 76.99 & 74.16 & 80.90 & \textbf{79.87} & 79.16 & 79.43 & 79.33 & 77.20 & 82.26 & 81.39 & 81.34 & 81.18 & \textbf{81.72} & 79.92 & 84.11 & \textbf{83.95} & 83.90 & 83.95 & 82.86 & 81.83 \\
WinoGrande & 68.11 & \textbf{67.01} & 65.67 & 66.30 & 66.14 & 62.43 & 74.35 & \textbf{72.93} & 72.69 & 72.85 & 72.30 & 69.46 & 74.98 & \textbf{75.45} & 73.16 & 74.21 & 74.51 & 73.32 & 81.93 & \textbf{80.43} & 80.27 & 78.30 & 79.72 & 73.64 \\
BoolQ & 77.28 & \textbf{75.11} & 74.07 & 72.39 & 74.34 & 69.69 & 82.97 & \textbf{81.19} & 80.46 & 79.11 & 81.07 & 71.65 & 85.32 & \textbf{86.30} & 85.47 & 85.89 & 85.63 & 84.46 & 87.16 & \textbf{86.36} & 86.27 & 85.08 & 86.06 & 83.30 \\
CSQA & 77.15 & \textbf{75.10} & 72.56 & 74.45 & 74.77 & 62.41 & 70.93 & \textbf{69.12} & 66.01 & 65.77 & 65.85 & 55.61 & 84.36 & \textbf{83.70} & 82.72 & 83.70 & 83.05 & 79.93 & 78.30 & \textbf{77.48} & 75.92 & 74.77 & 75.84 & 66.91 \\
Avg. & 70.27 & \textbf{68.62} & 67.37 & 67.38 & 67.83 & 63.05 & 73.93 & \textbf{72.30} & 71.29 & 70.74 & 71.61 & 65.61 & 78.19 & \textbf{77.95} & 77.09 & 77.45 & 77.14 & 75.12 & 79.70 & \textbf{79.15} & 78.75 & 77.89 & 78.31 & 73.81 \\
Recovery & 100\% & \textbf{77\%} & 60\% & 60\% & 66\% & 0\% & 100\% & \textbf{80\%} & 68\% & 62\% & 72\% & 0\% & 100\% & \textbf{92\%} & 64\% & 76\% & 66\% & 0\% & 100\% & \textbf{91\%} & 84\% & 69\% & 76\% & 0\% \\
\hline
\multicolumn{25}{l}{\textbf{MMLU (five-shot)}}\\
\hline
STEM & 61.85 & \textbf{58.10} & 56.74 & 56.74 & 56.42 & 49.67 & 56.45 & \textbf{53.31} & 52.46 & 52.87 & 52.14 & 45.35 & 77.70 & \textbf{76.44} & 75.96 & 75.64 & 75.64 & 70.31 & 71.17 & \textbf{69.20} & 68.51 & 68.82 & 68.25 & 56.14 \\
Hum. & 57.36 & \textbf{54.50} & 52.26 & 54.05 & 54.01 & 47.59 & 59.74 & \textbf{57.79} & 55.88 & 56.96 & 56.92 & 48.63 & 74.50 & \textbf{72.73} & 72.37 & 72.31 & 71.82 & 65.91 & 75.47 & \textbf{72.07} & 71.84 & 71.92 & 70.88 & 43.89 \\
Social & 76.93 & \textbf{73.29} & 70.59 & 71.82 & 72.50 & 63.05 & 76.37 & \textbf{73.58} & 71.40 & 71.40 & 72.54 & 63.70 & 87.16 & \textbf{86.35} & 85.77 & 85.96 & 85.28 & 82.26 & 88.07 & \textbf{86.84} & 86.25 & 86.38 & 86.35 & 74.26 \\
Other & 71.07 & \textbf{68.04} & 66.72 & 66.95 & 67.72 & 58.64 & 71.97 & \textbf{69.78} & 67.69 & 69.42 & 68.68 & 60.38 & 82.62 & \textbf{82.01} & 80.66 & 81.14 & 81.01 & 76.92 & 82.49 & 81.20 & 81.07 & 81.24 & \textbf{81.46} & 67.91 \\
Avg. & 65.69 & \textbf{62.42} & 60.48 & 61.40 & 61.64 & 53.89 & 65.35 & \textbf{62.90} & 61.12 & 61.96 & 61.88 & 53.80 & 79.79 & \textbf{78.60} & 77.94 & 78.00 & 77.66 & 72.92 & 78.82 & \textbf{76.68} & 76.29 & 76.45 & 76.02 & 58.61 \\
\emph{Recovery} & 100\% & \textbf{72\%} & 56\% & 64\% & 66\% & 0\% & 100\% & \textbf{79\%} & 63\% & 71\% & 70\% & 0\% & 100\% & \textbf{83\%} & 73\% & 74\% & 69\% & 0\% & 100\% & \textbf{89\%} & 87\% & 88\% & 86\% & 0\% \\
\Xhline{1.2pt}
\end{tabular}}
\end{table*}

\begin{table}[t]
\centering
\scriptsize
\renewcommand{\arraystretch}{1.12}
\setlength{\tabcolsep}{9pt}
\caption{WikiText-2 perplexity (lower is better).}
\label{tab:ppl}
\begin{tabular}{l|cccc}
\Xhline{1.2pt}
\textbf{Method}    & Qwen2.5        & Llama-3.1        & Qwen2.5        & Llama-3.1 \\
                   & 3B             & 8B              & 14B            & 70B \\
\hline
FP16               & 8.03           & 6.24            & 5.29           & 2.81 \\
\MXFPfour\ (bs=16) & 10.76          & 8.30            & 6.76           & 4.94 \\
\MMXFP\            & 9.11           & 6.97            & 5.96           & 3.62 \\
\NVFPfour\         & 8.95           & 6.94            & 5.93           & 3.63 \\
\MXFPfour\ (bs=32) & 11.03          & 8.31            & 6.94           & 4.88 \\
SMX4               & 468.5          & 11.86           & 9.15           & 9.43 \\
BlockDialect       & 9.30           & 7.13            & 6.07           & 3.79 \\
\AdaMXSixteen      & \textbf{8.80}  & \textbf{6.86}   & \textbf{5.84}  & \textbf{3.52} \\
\AdaMXThirtyTwo    & \textbf{9.17}  & \textbf{7.10}   & \textbf{6.04}  & \textbf{3.61} \\
\Xhline{1.2pt}
\end{tabular}
\end{table}

\section{Evaluation}
\label{sec:eval}

\subsection{Methodology}
\label{sec:eval:method}

\textbf{Models and benchmarks.}
We evaluate \AdaMX\ on four LLMs from two families,
spanning parameter scales from 3B to 70B:
Llama-3.1-8B and Llama-3.1-70B \cite{llama3},
and Qwen2.5-3B and Qwen2.5-14B \cite{qwen25},
together with the multimodal Gemma-4 12B \cite{gemma4}.
We report WikiText-2 perplexity \cite{wikitext}
with a sliding window at context length and stride of 2048 \cite{gptq, W4a8kv4},
and zero-shot accuracy with lm-evaluation-harness \cite{lmevalharness}
on six commonsense tasks:
ARC-Challenge \cite{arc},
HellaSwag \cite{hellaswag},
PIQA \cite{piqa},
WinoGrande \cite{winogrande},
BoolQ \cite{boolq},
and CommonsenseQA \cite{commonsenseqa}.
We also report 5-shot MMLU \cite{mmlu}
across its STEM, humanities, social sciences, and other categories.
For KV-cache quantization, we evaluate W4A4KV4 inference on Llama-3.1-8B using RULER \cite{ruler} at context lengths of 8K and 32K.

\textbf{Quantization.}
We apply \AdaMX\ to both weights and activations,
keeping the embedding layer in FP16.
At block size 16,
\MXFPfour\ and \AdaMXSixteen\ both use 4 element bits and an 8-bit shared field, giving 4.5 \EBW;
\AdaMX\ redistributes that field, encoding weights as \WFmt\ and activations as \AFmt\ (\cref{sec:twolevel}),
while doubling the block to 32 gives 4.25 \EBW.

\begin{table}[t]
\centering
\scriptsize
\renewcommand{\arraystretch}{1.12}
\setlength{\tabcolsep}{4pt}
\caption{Multimodal accuracy (\%) for Gemma-4 12B.
}
\label{tab:multimodal}
\resizebox{1\columnwidth}{!}{
\begin{tabular}{l|ccccc}
\Xhline{1.2pt}
\textbf{Benchmark} & FP16 & \AdaMXSixteen & \MXFPfour-16 & \AdaMXThirtyTwo & \MXFPfour-32 \\
\hline
MMMU                          & 60.8 & \textbf{58.3} & 48.9 & 54.2 & 42.2 \\
\quad Art \& Design           & 67.5 & \textbf{63.3} & 59.2 & 60.0 & 49.2 \\
\quad Business                & 65.7 & \textbf{66.4} & 52.2 & 58.2 & 43.3 \\
\quad Science                 & 57.7 & \textbf{56.2} & 44.5 & 44.5 & 40.9 \\
\quad Health \& Medicine      & 59.3 & \textbf{56.6} & 45.5 & 54.5 & 37.9 \\
\quad Humanities \& Soc.\ Sci. & 72.3 & \textbf{69.7} & 63.0 & 67.2 & 48.7 \\
\quad Tech \& Engineering     & 49.5 & \textbf{49.0} & 37.0 & 44.8 & 37.0 \\
\hline
ScienceQA                     & 82.8 & \textbf{79.6} & 74.3 & 76.3 & 67.4 \\
AI2D                          & 77.7 & \textbf{67.9} & 61.6 & 59.5 & 45.7 \\
TextVQA                       & 67.7 & \textbf{60.7} & 55.8 & 59.1 & 51.5 \\
\hline
Avg.                          & 72.3 & \textbf{66.6} & 60.2 & 62.3 & 51.7 \\
Recovery               & 100\% & \textbf{72\%} & 41\% & 51\% & 0\% \\
\Xhline{1.2pt}
\end{tabular}}
\end{table}

\textbf{Baselines and implementation.}
We compare \AdaMX\ against FP16 and other low-bit formats:
\MXFPfour\ \cite{ocpmx, rouhani2023mx} at block size 16 (4.5 \EBW) and 32 (4.25 \EBW),
\MMXFP\ \cite{m2xfp} at its reported block size of 32 (4.5 \EBW),
and \NVFPfour\ \cite{nvfp4}.
\MMXFP\ and \NVFPfour\ are evaluated using publicly available implementations.
\AdaMXSixteen\ is matched against \MMXFP, \NVFPfour, and \MXFPfour\ at 4.5 \EBW,
and \AdaMXThirtyTwo\ against \MXFPfour\ at 4.25 \EBW.
Our quantization framework extends the Microsoft microxcaling library \cite{rouhani2023mx} in PyTorch \cite{pytorch},
and all accuracy results are measured on a single server with four NVIDIA H100 80GB GPUs \cite{h100}.

\textbf{Hardware implementation.}
We implement the \AdaMX\ decode units, processing elements, and quantization logic in Verilog.
The design is synthesized and placed and routed using a 22-nm FD-SOI standard-cell library at 500 MHz and 0.8~V.
\cref{sec:eval:hw} reports the resulting area and power.
For a fair comparison,
the \AdaMX\ and \MXFPfour\ accelerators are matched in throughput and buffer size,
so the area difference reflects the decode and correction logic that \AdaMX\ adds.

\textbf{Hardware baselines.}
The hardware comparison is against an \MXFPfour\ and \NVFPfour processing element synthesized in the same flow using the same architecture,
which isolates the cost of the \AdaMX logic.

\subsection{Main Results}
\label{sec:eval:main}
\AdaMX\ provides two operating points that share one encoding and differ only in block size:
\AdaMXSixteen\ at 4.5 \EBW and \AdaMXThirtyTwo\ at 4.25 \EBW.
Each point is evaluated against the baselines at its own \EBW.

\textbf{Perplexity.}
\cref{tab:ppl} reports WikiText-2 perplexity, where lower is better.
\AdaMXSixteen\ reaches the lowest perplexity on every model,
narrowing the \MXFPfour\ gap to FP16 by 63 to 72 percent,
and \AdaMXThirtyTwo\ narrows it by 55 to 62 percent at the lower budget.

\textbf{Downstream accuracy.}
\cref{tab:downstream} reports zero-shot commonsense accuracy across six tasks
and 5-shot MMLU across four categories, where higher is better.
\AdaMXSixteen\ retains at least 97 percent of the FP16 commonsense average on every model,
and at least 95 percent of the MMLU category average,
with the gap to FP16 decreasing for the 14B and 70B models.
At the same 4.5 \EBW, it leads \NVFPfour\ on both benchmarks,
while at the lower 4.25 \EBW, \AdaMXThirtyTwo\ remains ahead of \MXFPfour\ and performs comparably to \NVFPfour.
\MXFPfour\ degrades far more on MMLU than on the commonsense tasks,
suggesting that these knowledge-intensive tasks are more sensitive to its coarse representation.
The advantage is consistent across both the Llama and Qwen families and from 3B to 70B parameters,
showing that the gains persist across model families and parameter scales.

\begin{table}[t]
\centering
\scriptsize
\renewcommand{\arraystretch}{1.3}
\setlength{\tabcolsep}{3pt}
\caption{RULER accuracy (\%) on Llama-3.1-8B under W4A4KV4 inference at 8K and 32K context. Each row averages the RULER task variants in that group, counted in parentheses.}
\label{tab:ruler_kv}
\resizebox{1\columnwidth}{!}{
\begin{tabular}{l|ccccc|ccccc}
\Xhline{1.2pt}
& \multicolumn{5}{c|}{8K Context} & \multicolumn{5}{c}{32K Context} \\
\cline{2-6}\cline{7-11}
Task & FP16 & \AdaMXSixteen & \NVFPfour & \AdaMXThirtyTwo & \MXFPfour & FP16 & \AdaMXSixteen & \NVFPfour & \AdaMXThirtyTwo & \MXFPfour \\
\hline
NIAH single (3)            & 100.0 & \textbf{99.8} & 99.3 & 99.3 & 87.8 & 100.0 & \textbf{98.5} & \textbf{98.5} & 97.2 & 69.2 \\
NIAH multi-key (3)         & 99.2 & \textbf{96.0} & 86.5 & 84.8 & 33.7 & 97.7 & \textbf{84.0} & 62.3 & 65.3 & 18.5 \\
NIAH multi-value/query (2) & 99.9 & \textbf{96.7} & 96.6 & 94.9 & 38.8 & 99.3 & \textbf{94.1} & 91.3 & 85.6 & 19.3 \\
Variable tracking (1)      & 99.7 & \textbf{99.3} & 97.8 & 96.8 & 67.7 & 99.5 & \textbf{95.9} & 93.6 & 91.9 & 40.7 \\
Word extraction (2)        & 93.6 & 85.9 & \textbf{87.6} & 81.4 & 28.8 & 75.4 & \textbf{67.5} & 61.8 & 54.4 & 17.1 \\
Question answering (2)     & 56.3 & \textbf{56.8} & 55.8 & 49.0 & 27.0 & 52.5 & 47.8 & \textbf{50.5} & 44.5 & 18.5 \\
\hline
\textbf{Avg. (13)}         & 92.0 & \textbf{89.7} & 87.3 & 84.6 & 47.8 & 88.2 & \textbf{81.7} & 75.6 & 72.9 & 31.8 \\
\Xhline{1.2pt}
\end{tabular}}
\end{table}

\textbf{Multimodal accuracy.}
\cref{tab:multimodal} reports the accuracy of \AdaMX\ on the vision-language model Gemma-4 12B~\cite{gemma4}
across four image benchmarks:
MMMU \cite{mmmu}, ScienceQA \cite{scienceqa}, AI2D \cite{ai2d}, and TextVQA \cite{textvqa}.
\AdaMXSixteen\ retains 87 to 96 percent of the FP16 accuracy across the four benchmarks,
with the largest drop on AI2D.
At each memory budget \AdaMX\ stays ahead of \MXFPfour\ on every benchmark.
On MMMU the margin reaches 12 points.
These results show that \AdaMX's accuracy improvements extend to multimodal inference.

\textbf{Long-context KV cache.}
We also measure the performance of the KV cache quantization in long-context scenarios.
Each K/V entry is written once and reused by all subsequent decoding steps,
giving the KV cache a read-mostly reuse pattern closer to weights than to transient activations.
We therefore encode it with \AdaMX's weight-side format and apply the same per-block format-selection rule.
Although this per-block selection incurs a one-time encoding cost when the block is written, the cost is amortized over its repeated reads in subsequent decoding steps.
We evaluate W4A4KV4 inference on Llama-3.1-8B using RULER \cite{ruler} at context lengths of 8K and 32K (\cref{tab:ruler_kv}).
At the matched 4.5 \EBW\ point, \AdaMXSixteen\ exceeds \NVFPfour\ by 2.4 points at 8K, and the margin widens to 6.1 points at 32K on the 13-task average.
\AdaMXSixteen\ ranks first among the low-bit formats on five of the six task groups at each length.
At 4.25 \EBW, \AdaMXThirtyTwo\ also outperforms \MXFPfour\ by 36.8 and 41.1 points at 8K and 32K, respectively.

\subsection{Hardware Evaluation}
\label{sec:eval:hw}

\textbf{Area and power.}
\cref{tab:area} reports area and power for the \AdaMX\ accelerator and its processing element, with \MXFPfour\ as the baseline.
Both designs use a 2$\times$2 MAC PE, contain the same number of multipliers (8192), and share an identical 864~KB SRAM buffer, so the comparison is matched in throughput and memory.
The \MXFPfour\ array totals 2.10~mm$^2$ and the \AdaMX\ array 2.16~mm$^2$, a 3.2\% increase.
This increase comes from the block-max extension and scale refinement that \AdaMX\ adds in each PE.
The power follows the same pattern: at 500~MHz the \AdaMX\ array draws 482~mW and the \MXFPfour\ array 466~mW, a 3.5\% increase.
The \AdaMX\ MAC array sustains a peak 8.19~TOPS at 500~MHz, corresponding to a compute efficiency of 17.1 TOPS/W.

\begin{figure}[t]
\centering
\includegraphics[width=\columnwidth]{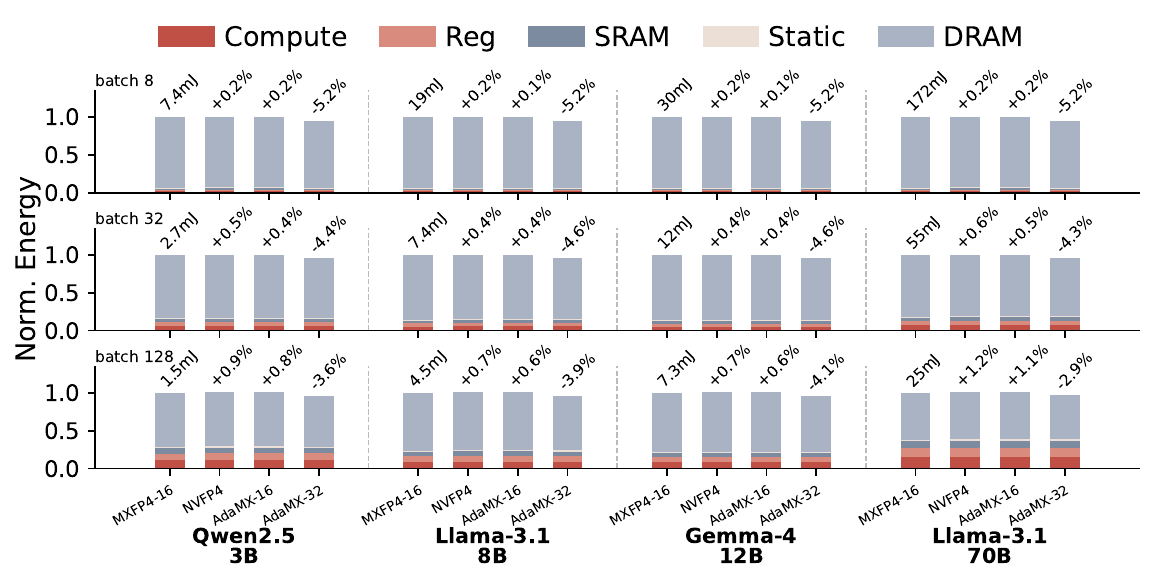}
\caption{Decode energy across four models and three batch sizes, normalized to \MXFPfour\ at block size 16.}
\label{fig:energy}
\end{figure}

\textbf{Energy breakdown.}
We model end-to-end inference with SCALE-Sim~\cite{scalesimv3}.
The on-chip buffer is 864~KB and the clock is 500~MHz.
Per-MAC energy is obtained from PrimeTime power analysis, SRAM energy from Accelergy~\cite{accelergy} and data sheets, and DRAM energy from CACTI~\cite{cacti7}.
All four formats share the array size, clock, and buffer, so they run at identical throughput.
The workload is decode at batch 8, 32, and 128, over the 2048-token context.
\cref{fig:energy} reports decode energy for four models, normalized to the \MXFPfour\ block-16 total of each model.
External memory dominates at 61 to 94 percent of system energy.
\AdaMXSixteen\ raises the per-MAC energy by 2.9 percent over \MXFPfour, which adds only 0.1 to 1.1 percent to system energy.
This modest energy overhead accompanies the substantial accuracy improvements reported in \cref{tab:ppl} and \cref{tab:downstream}.
Relative to \MXFPfour\ at block size 16, \AdaMXThirtyTwo\ reduces system energy by 2.9 to 5.2 percent.
\AdaMXThirtyTwo\ trades some accuracy relative to \AdaMXSixteen, but it is still more accurate than \MXFPfour\ at the same 4.25 \EBW.
This reduction follows from the lower \EBW: block size 32 spends 4.25 bits per element rather than 4.5 at block size 16, so each weight access transfers fewer bits.
The block-size saving therefore outweighs the cost of the decode logic.
At matched throughput and memory, \AdaMX\ closes 55 to 72 percent of the \MXFPfour\ perplexity gap to FP16 (\cref{tab:ppl}), at 3.2 percent more area and 3.5 percent more power in the compute array, and at most 1.1 percent more system energy.

\begin{table}[t]
\centering
\small
\renewcommand{\arraystretch}{1.12}
\setlength{\tabcolsep}{4pt}
\caption{Area and power of \MXFPfour\ and \AdaMX\ (22nm FDSOI technology, 500 MHz).}
\label{tab:area}
\resizebox{\columnwidth}{!}{
\begin{tabular}{l l r r}
\Xhline{1.2pt}
\textbf{Component}                & \textbf{Role}                    & \textbf{Area ($\mu$m$^2$)} & \textbf{Power (mW)} \\
\hline
\MXFPfour\ SA \textit{(baseline)} & compute $+$ 864\,KB SRAM         & 2099260.8 & 465.78 \\
\MXFPfour\ PE                     & Same throughput as \AdaMX\ PE        & 6992.1 & 5.92 \\
\hline
\AdaMX\ SA                        & compute $+$ 864\,KB SRAM         & \textbf{2166162.5}~{\footnotesize($+3.2\%$)} & \textbf{482.05} \\
\quad SRAM                        & 864\,KB, on-chip storage                  & 1645925.0 & 81.04 \\
\quad PE array ($\times 64$)      & compute                          & 509504.4 & 392.05 \\
\quad Decoders ($\times 32$)      & weight \& activation decode      & 6050.1 & 4.01 \\
\quad Quantizers ($\times 4$)     & activation re-encode             & 3226.0 & 1.52 \\
\quad Loader $+$ glue             & streaming $+$ interconnect       & 1457.1 & 3.42 \\
\AdaMX\ PE                        & per-PE logic, 4$\times$ 32-lane MAC & \textbf{7961.0} & \textbf{6.13} \\
\quad MAC array ($\times 4$)      &                                  & 6168.4 & \\
\quad\quad per MAC ($\times 1$)   &                                  & 1542.1 & \\
\quad\quad\quad Adder tree        & accumulation                     & 578.5 & \\
\quad\quad\quad Emul              & element mul $+$ align            & 367.6 & \\
\quad\quad\quad \AdaMX\ enhancement & block-max ext.\ $+$ scale refine. & 270.6 & \\
\quad\quad\quad Other \& regs.    & incl.\ FP32 normalize             & 325.4 & \\
\quad Other \& regs.\ (top)       & decoded-operand regs $+$ FP32 acc $+$ glue & 1792.6 & \\
\Xhline{1.2pt}
\end{tabular}
}
\end{table}

\section{Conclusion}
\label{sec:conclusion}

Quantization heterogeneity arises across both blocks and operands.
We introduce \AdaMX, which repurposes MX's over-provisioned exponent bits as operand-specialized metadata: weights route each block among four format-enhancement modes, 
while activations use round-to-nearest E4M1 scaling with a lossless \FPsix block-maximum extension. 
The same weight-oriented format is further extended to KV-cache quantization.
We design a 22-nm FD-SOI accelerator that supports block sizes 16 and 32 on a unified datapath while efficiently decoding the proposed metadata.
Across LLMs from 3B to 70B and the Gemma-4 12B multimodal model,
\AdaMX substantially outperforms \MXFPfour and exceeds \NVFPfour at comparable bit budgets. \AdaMXSixteen adds at most
$1.1\%$ system energy, 
while \AdaMXThirtyTwo reduces it by
$2.9$--$5.2\%$. 
These results demonstrate that lightweight,
operand-aware metadata can improve the accuracy--efficiency tradeoff of 4-bit microscaling with only modest additional hardware complexity.

\section*{Acknowledgment}
The authors thank Google Cloud for providing compute resources, Google Research for supporting this work, and GlobalFoundries and its University Partnership Program for technology access.

\bibliographystyle{IEEEtranS}
\bibliography{refs}

\end{document}